# Emission of photon pairs in optical fiber - effect of zero-point fluctuations

V. Hizhnyakov

*Institute of Physics, University of Tartu, W. Ostwaldi 1, 50411 Tartu, Estonia*
hizh@ut.ee

**Abstract.** A nonlinear quantum-optical process is considered - emission of photon pairs by the reflecting end of a fiber excited by a standing laser wave. Radiation occurs due to periodic changes in the optical length of the fiber over time. This radiation can be significantly enhanced in long fibers. Because of this enhancement, a nonperturbative description of the process is required. Such a description presented here predicts a strong peak of radiation, if the amplitude of the oscillations of the optical length coincides with half of the laser wavelength. The considered phenomenon can be used to create a source of entangled photons.

## I. INTRODUCTION

Good optical fibers may have very small losses $\lesssim 0.2$ dB/km. Due to this property, they are of great interest for nonlinear optics, since they allow one to study weak processes, which in other cases are difficult to observe. Here we are considering one such process – the spontaneous emission of the pairs of entangled photons by the reflecting end of a fiber excited by a standing laser wave. The radiation under consideration is generated by a periodic change of the fiber refractive index $n$ caused by the Kerr nonlinearity. This change, in turn, leads to a periodic change in the optical length of the fiber over time and to similar changes in the zero-point state of fiber modes. The mentioned changes lead to excitation of modes and appearance of photons. Although the amplitude $\delta n$ of the variation in time of the refractive index is very small (10-8 ÷ 10-10), the amplitude of the change in the optical length of fiber can be quite large compared to the optical wavelength, allowing one to expect that the arising radiation can be detected.

The radiation under consideration has a close analogy with the emission of the photon pairs caused by the time variation of the length of the resonator(s) or the refractive index of the medium (see, e.g. [1-9]). The latter emission is usually called the dynamical Casimir effect (DCE) [2,4]. DCE directly reveals the properties of the quantum vacuum state. Therefore, it is not surprising that this effect has been considered in a large number of works and in several books (see, e.g. the books [10-17]). It was found that under normal conditions this effect is extremely weak; it is not negligible only when the rate of the change of the optical length of the resonator with time approaches the speed of light $c$ [1]. The main attention has been on DCE in small-size cavities, which are comparable to the wavelength of DCE radiation (see [16] and references therein). In this case, the parametric resonance of one of the resonator modes with external perturbation can be achieved. This allows one to significantly increase the energy of resonant modes by periodic perturbation and to generate many photons, despite of a small number of working modes. Such enhanced DCE was observed in a microwave range using superconducting circuits [17].

In the optical range, due to the smallness of wavelength, there is another opportunity for DCE amplification – to excite the oscillation of the optical length in a long fiber with one transparent and one reflective end. The fiber can be excited through the transparent end. Photons will be generated by the reflective end and will be emitted back through the transparent end along with the reflected photons and distinguished from the reflected photons based on their spectrum. It should be noted that this radiation has a close analogy with the Fulling-Davis-Unruh effect [18,19] – the spontaneous emission of photon pairs by an accelerating mirror. In both cases, a continuous spectrum of electromagnetic modes operates; one can generate a significant (even large) number of photons, so that at any time in each mode there will be an infinitely small probability of generation of photons. This differs essentially from the DCE in small resonators, where a large number of photons must be generated in resonant mode(s).

Another significant feature of the considered radiation, which distinguishes it from ordinary nonlinear optical processes, is the absence of a phase matching condition. Instead, it is necessary here that the refractive index oscillates in the phase through the length of the fiber.

We take into account that a standing laser wave causes the following change in the optical length of the fiber:

$$\delta L_t = 4n_2 \cos^2(\omega_0 t) \int_0^L dx I(x) \cos^2(\omega_0 x/c).$$

Here $x$ is the coordinate along the fiber, $n_2$ is its nonlinear refractive index, $I(x)$ is the intensity of laser light, $\omega_0$ is its frequency, $L$ is the length of the standing laser wave. After integration we get $\delta L_t = 2\delta n L \cos^2(\omega_0 t)$, where $\delta n = n_2 I$.

It turns out that if the double (full) amplitude of the change in the optical fiber length $2a = 2\delta n L$ reaches the half-wavelength of the laser light $\lambda_0/2 = \pi c/\omega_0$ (that corresponds to appearance of an additional mode) then the intensity of the two-photon radiation acquires a peak. In this case the mean square velocity of the optical length is comparable to $c$. This amplitude is obtained when the laser power is $W \sim S\lambda_0/2n_2 L$, where $S$ is the cross-section area of the fiber. In the optical range $\lambda_0 \sim 0.5\mu$ silica fibers have $n_2 \approx 3.5 \cdot 10^{-20}$ m2/W. For a fiber with $L \sim 100$ m and $S \sim 10^{-10}$ m$^2$, it turns out that $2a \sim \lambda_0/2$, if the standing laser wave has a duration $L/c \sim 0.3\mu s$ and energy $\sim 1\mu J$.

The presented estimate shows that fiber optics is well suited for observing the two-photon radiation in the optical range caused by a change in the zero-point state with time. Moreover, in the case of excitation with $2a \sim \lambda_0/2$, the changes in the zero-point state of the fiber modes are not weak and, therefore, the generation should be considered nonperturbatively. The required description can be given by using the theory of quantum vacuum effects in strong fields; see, for example the books [19,20]. Here we apply this theory, adapted in [21, 22] to describe a similar process – the two-phonon decay of strong vibrations in solids.

Entangled photon pairs can be generated in fibers by using spontaneous parametric down conversion (SPDC) for the modes that satisfy the phase matching condition [23]. However, this condition is difficult to fulfill at long distances, which limits the effectiveness of SPDC in fibers.

## II. GENERATION OF PHOTON PAIRS

According to the theory [19] (see also [20,21]), the time-dependent classical field changes the positive-frequency-type time dependence of the destruction operator $\hat{b}_k$ of the quantum field adding also a term with the negative-frequency time dependence, corresponding to the initial creation operator. As a result, field operators undergo the Bogolyubov-type transformations; and the number of particles generated by the classical field is determined by the terms with the changed time-dependence [19-24].

To find the number of photons generated in a fiber, let us consider a nonlinear one-dimensional wave equation for the field potential operator $\hat{A}$, taking into account the Kerr nonlinearity. In the case $|\delta n_t| \ll 1$ this equation gets the form

$$\ddot{\hat{A}} - (c/n_t)^2 \hat{\nabla}^2 \hat{A} = 0 \tag{1}$$

Let us first suppose that both ends of the fiber are reflective. We consider the solution $\hat{A} = \sum_k \hat{A}_k(t) \sin(\pi k x/L_t)$, satisfying the boundary condition $\hat{A} = 0$ at $x = 0$ and at $x = L_t$, $\hat{A}_k(t) = \sqrt{\hbar/2\omega_k \varepsilon_0 L} \left( \hat{b}_k(t) + \hat{b}_k^+(t) \right)$ is the field operator of the mode $k$ with the frequency $\omega_k = ck/L$, $\varepsilon_0$ is the vacuum permittivity. Inserting this solution into the wave equation (1) we get in the large $L$ limit

$$\sum_k (\ddot{\hat{A}}_k + \omega_k^2 \hat{A}_k) \sin(ky) = \frac{y}{L} \sum_{k'} k'(2\dot{L}_t \dot{\hat{A}}_{k'} + \ddot{L}_t \hat{A}_{k'}) \cos(k'y),$$

where $y = \pi x/L$. As a next step, we apply the saw wave equation $y = -\sum_{j \geq 1} (-1)^j j^{-1} \sin(jy), |y| \leq \pi$ and take into account that only the terms $j = k \pm k'$ contribute to the term $k$. This leads to the replacement $j^{-1} \to k/(k^2 - k'^2)$. We restrict our consideration to the large time limit $t \to \infty$. In this limit only the terms with $\omega_k + \omega_{k'} = 2\omega_0$ are important, giving $2\dot{L}_t \dot{\hat{A}}_{k'} + \ddot{L}_t \hat{A}_{k'} \cong 2a\omega_0(2\omega_k - 2\omega_0)\hat{A}_{k'}$. As a result, the factor $(k^2 - k'^2)^{-1}$ is cancelled, and one gets the following equation for the operator $\hat{A}_k = (-1)^k \left( \hat{b}_k + \hat{b}_k^+ \right)$:

$$\ddot{\hat{A}}_k + \omega_k^2 \hat{A}_k \cong \omega_k \cos(2\omega_0 t) \hat{\Lambda}(t), \tag{2}$$

where $\hat{\Lambda} = (2a/L) \sum_k \omega_k \hat{A}_k$. The right-hand term of this equation describes the perturbation of the mode $k$, stemming from the oscillations of the optical length in time. This term is factorized, which is a decisive simplification of the equations for the field operators in the limit of large $L$ and $t$, as compared to similar equations for a cavity with an arbitrary $L$ and $t$ [1,16]. The consequence of factorization is the localization of perturbation in space: the sum $\sum_{k'} \omega_{k'} \hat{A}_{k'} \sin(\pi k x n_t/L)$ differs significantly from zero only for $x \sim \lambda_0$ and $x \sim Ln - \lambda_0$. Therefore, the reflective ends emit photons independently, and their number is twice as much as in the case of a single reflective end which is that we are in fact interested in.

We present now Eq.(2) in the integral form

$$\hat{A}_k(t) = \hat{A}_k^{(0)}(t) + \frac{2a}{L} \int_0^t dt_1 \sin(\omega_k(t - t_1)) \cos(2\omega_0 t_1) \hat{\Lambda}(t_1) \tag{3}$$

where $\hat{A}_k^{(0)}$ is the initial value of $\hat{A}_k$.

Weak excitation. If laser intensity $I$ is small then one can take $\hat{\Lambda} \cong \hat{\Lambda}_0 = (2a/L) \sum_k \omega_k \hat{A}_k^{(0)}$. In this case, for large $t$, the positive frequency field operator gets the form

$$\hat{b}_k(t) = \left( \hat{b}_k^{(0)} + \sum_{k'} v_{kk'} \left( \hat{b}_{k'}^{(0)+} - \hat{b}_{k'}^{(0)} \right) \right) e^{-i\omega t}, \tag{4}$$

where $v_{kk'} = i(a/2L) e^{i\phi_{k'}} \sqrt{\omega_k \omega_{k'}} \int_0^t e^{i(2\omega_0 - \omega_k - \omega_{k'})t_1} dt_1$, $\phi_{k'}$ is the random phase (sum over $k'$ includes averaging over phases $\phi_{k'}$). Due to the positive-frequency time dependence, this operator should be treated as the final destruction operator [19,20]. Thus, the oscillations of $L$ in time cause the transformation of the destruction operator into the sum of the initial destruction and creation operators. The zero-point state is not the zero state of the final destruction operators, which means that photons appear in the fiber. The number of generated photons with the frequency $\omega_k$ equals $n_k = \sum_{k'} |V_{kk'}|^2$, giving for large $t$ the constant rate of generation $\dot{n}_k = \pi a^2 \omega_k (2\omega_0 - \omega_k)/2L^2$. Replacing the sum over $k$ by the integral over frequency $\omega$ one finds the following spectral rate of generated photons:

$$\dot{N}_0(\omega) \simeq \pi \upsilon^2 \omega (2\omega_0 - \omega)/2\omega_0^2, \tag{5}$$

where $\upsilon = \omega_0 a/c = \omega_0 n_2 I/c$ is the amplitude of the change in the dimensionless rate of the optical length.

Arbitrary excitation. To find the number of photons generated for an excitation with arbitrary intensity $I$, let us consider the equation for $\hat{b}_k = (-1)^k \hat{b}_k$:

$$\hat{b}_k(t) = \hat{b}_k^{(0)} e^{-i\omega_k t} + \int_0^t dt_1 \sin(\omega_k(t-t_1)) \cos(2\omega_0 t_1) \hat{\Lambda}^+(t_1), \tag{6}$$

which follows from Eq. (3). Here $\hat{\Lambda}^+(t)$ is the positive frequency part of $\Lambda(t)$, which satisfies the equation

$$\hat{\Lambda}^+(t) = \hat{\Lambda}_0^+(t) + \frac{a^2}{L^2} \int_0^\infty dt_1 \int_0^\infty dt_2 G(t-t_1) G(t_1-t_2) e^{2i\omega_0(t_1-t_2)} \hat{\Lambda}^+(t_2).$$

Here $G(t) = \Theta(t) \sum_k \omega_k \sin \omega_k t$, $\Theta(t)$ is the Heaviside step function, $\hat{\Lambda}_0^+(t)$ is the positive frequency part of $\hat{\Lambda}_0(t)$ (the first-order term with respect to $a$ does not contribute to $\hat{\Lambda}^+(t)$; fast oscillating terms are neglected). Using spectral representation of the last equation, we get a simple linear equation [25]. Solving this equation, we find [21, 22]

$$\hat{\Lambda}^+(t) = cL^{-1} \sum_k \hat{b}_k^{(0)} \sqrt{\omega_k} e^{-i\omega_k t} / (1 - \upsilon^2 R(\omega_k)) \tag{7}$$

where

$$R(\omega) = G(\omega/\omega_0) G^*(2 - \omega/\omega_0),$$
$$G(x) = \left(2 + x \ln((2-x)/(2+x)) + i\pi x/2\right)/2\pi.$$

This gives

$$\dot{N}(\omega) \simeq \frac{\dot{N}_0(\omega)}{|1 - \upsilon^2 R(\omega)|^2}. \tag{8}$$

If the excitation is performed by long pulse with duration, $\Gamma^{-1} = L/c$, then the spectral rate of generated photons with the frequency $\omega$ equals $N(\omega) \cong \Gamma^{-1} \dot{N}(\omega)$.

## III. DISCUSSION

In the case of weak excitation the total number of emitted photons by single reflecting fiber end equals

$$\mathrm{N} = \pi^3 (4L/\lambda_0)^2 n_2^2 I^2 \omega_0 / 3\Gamma. \qquad (9)$$

The spectrum is broad: $N(\omega) \propto \omega(2\omega_0 - \omega)$, $0 \leq \omega \leq 2\omega_0$. Note a very fast (cubic for $\Gamma = c/L$) increase in radiation from $L$. That is why fiber optics is well suited for observing spontaneous two-photon radiation in the optical range caused by a change in the zero point state with time. The yield of emission is

$$\eta = \mathrm{N}/\mathrm{N}_0 \approx \pi^3 (4L/\lambda_0)^2 n_2^2 I \hbar \omega_0^2 / 3S, \qquad (10)$$

where $\mathrm{N}_0 = S \cdot I / \Gamma \hbar \omega_0$ is the number of excitation quanta. For silica fibers with $L = 100\,\mathrm{m}$, $S = 10^{-10}\,\mathrm{m}^2$, for laser with $\lambda_0 = 0.5\,\mu$ and $I = 10^6\,\mathrm{W/m}^2$ we get $\upsilon \sim 0.01$ and $\eta \sim 10^{-8}$. This yield, although small, strongly exceeds the typical generation yield $\eta_0 \sim 10^{-12}$ of photon pairs by spontaneous parametric down conversion (SPDC) in semiconductors and in stratified structures [26-28]. However, differently from SPDC in our case the emission has a wide spectrum (5).

The generated radiation increases with increasing laser intensity $I$ and rate $\upsilon$, and, according to Eq. (8) it becomes monochromatic if $\upsilon$ approaches $\upsilon_0 = 2.944$. However, if the intensity $I$ becomes so high that $\upsilon \gg \upsilon_0$, then $N \propto \upsilon^{-2} \propto I^{-2}$, that is, the intensity of the radiation quickly decreases with increasing $\upsilon$ and $I$, and its spectrum becomes broad again. Such an inverse dependence of radiation on $\upsilon$ was expected: if the optical length fluctuates faster than the zero-point fluctuations of modes, then the modes cannot perceive such oscillations, as a result of which the influence of the oscillations weakens. This type of inverse dependence has been experimentally observed for two-phonon relaxation of quasi-molecule $\mathrm{Xe}_2^*$ in experiments on hot luminescence in $\mathrm{Xe}$ crystals [29].

Note that the value of $\upsilon_0$ is close to $\pi$, and the full amplitude of oscillations of the optical length is close to $\lambda_0/2$, i.e. the full amplitude acquires a value that requires a periodic appearance and disappearance of the mode $\lambda_0$ along with its zero-point energy. There is an optimal pulse that allows one to get the maximum possible number of generated photon pairs. Such an optimal pulse has the intensity $I_0 \sim \lambda_0/2n_2 L$ and energy $\mathrm{E}_0 \sim \lambda_0 S/2n_2 c$. Taking $n_2 = 3.5 \cdot 10^{-20}\,\mathrm{m2/W}$, $\lambda_0 S \sim 10^{-17}\,\mathrm{m}^3$, we get $\mathrm{E}_0 \sim 10\,\mu\mathrm{J}$.

For $\upsilon$ close to $\upsilon_0$, the spectrum of generated photons is almost Lorentzian with a small width $\varsigma\Gamma + |\upsilon - \upsilon_0|\omega_0$ and a peak at $\omega_0$. The maximum number of generated photons is $\mathrm{N}_{max} \simeq 9\pi^4 \omega_0^2 / 4\upsilon_0^2 \Gamma^2$, supposing that $\mathrm{N}_{max} < \mathrm{N}_0$. In this case one can get very high yield $\eta \sim 1$. However, to obtain such a yield, it is necessary to control the intensity $I$ of the excitation with unrealistically high accuracy $\Gamma/\omega \sim 10^{-8}$. If $I$ is controlled to ten percent, then one gets $\mathrm{N} \sim 10^3\, L/\lambda_0$. In this case, the spectral width of the generated photons is $\delta\omega \approx 0.11\omega_0$. For fibers with $L \sim 100\,\mathrm{m}$ and $\lambda_0 \sim 0.5$ μm one gets $\mathrm{N} \sim 10^{12}$ and $\eta \sim 10^{-4}$.

## IV. CONCLUSIONS

In this communication the generation of pairs of entangled photons in a long optical fiber with a reflective end, excited by a standing laser wave, was considered. We found that high generation efficiency can be achieved. We predict a sharp peak of radiation for a laser pulse with an energy of $\sim 1\,\mu\mathrm{W}$, which is caused by a periodic in time change in the optical length of the fiber with an amplitude of $\lambda_0/2$. This amplitude corresponds to the creation and destruction of an additional mode in the fiber during each oscillation period. The spectrum of generated photons of the peak has a maximum at the frequency $\omega_0$ of the laser wave. The number of generated photons can be $N \sim 10^3 L/\lambda_0$ or more, which makes it possible to achieve a yield $\eta \gtrsim 10^{-4}$ in a fiber $\gtrsim 100$ m long. Consequently, in a long fiber one can turn a remarkable part of monochromatic photons into pairs of entangled photons with a wider spectrum.

Finally, we note that the nonlinear quantum-optical process discussed here may be of practical interest as a possible source of entangled photons. The coincidence of the mean radiation frequency with the laser frequency $\omega_0$ can be avoided if bi-chromatic excitation is used with the frequencies $\omega_1$ and $\omega_2$; in this case, the maximum frequency of two-photon radiation will be $(\omega_1 + \omega_2)/2$.

## ACKNOWLEDGEMENTS

The work was supported by Estonian Research Council grant PRG347.